\documentclass[fleqn,10pt]{wlscirep}
\usepackage[utf8]{inputenc}
\usepackage[T1]{fontenc}
\usepackage{bm,amsmath}
\usepackage{hyperref}
\usepackage{caption}
\usepackage{subcaption}
\usepackage{soul}
\title{Linear Laws of Markov Chains with an Application for Anomaly Detection in Bitcoin Prices}

\author[1,*,+]{Marcell T. Kurbucz}
\author[1,+]{P\'eter P\'osfay}
\author[1,+]{Antal Jakov\'ac}

\affil[1]{Wigner Research Centre for Physics, Department of Computational Sciences, 29-33 Konkoly-Thege Mikl\'os Street, Budapest, H-1121, Hungary}
\affil[*]{kurbucz.marcell@wigner.hu}
\affil[+]{These authors contributed equally to this work.}


\begin{abstract}
The goals of this paper are twofold: (1) %to present a new method that is able to find linear laws in Markov chains 
to present a new method that is able to find linear laws governing the time evolution of Markov chains and (2) to apply this method for anomaly detection in Bitcoin prices. To accomplish these goals, first, the linear laws of Markov chains are derived by using the time embedding of their (categorical) autocorrelation function. Then, a binary %Markov
series is generated from the first difference of Bitcoin exchange rate (against the United States Dollar). Finally, the minimum number of parameters describing the linear laws of this series is identified through stepped time windows. Based on the results, linear laws typically became more complex (containing an additional third parameter that indicates hidden Markov property) in two periods: before the crash of cryptocurrency markets inducted by the COVID-19 pandemic (12 March 2020), and before the record-breaking surge in the price of Bitcoin (Q4 2020 -- Q1 2021). In addition, the locally high values of this third parameter are often related to short-term price peaks, which suggests price manipulation.
\end{abstract}
\begin{document}

\flushbottom
\maketitle
\thispagestyle{empty}

%\noindent Please note: Abbreviations should be introduced at the first mention in the main text – no abbreviations lists. Suggested structure of main text (not enforced) is provided below.

\section*{Introduction}

The hidden Markov model \cite{baum1966statistical,baum1970maximization,baum1972inequality,rabiner1989tutorial} is a stochastic approach that captures hidden information from observable random variables. As an extension of Markov models, it can be characterized by two stochastic processes: an unobserved %, finite-state
Markov chain and an observed sequence that depends only on the current state of the aforementioned Markovian (memoryless) process \cite{danisman2021hidden}. Its objective is to estimate the hidden sequence of states that provides the best statistical explanation of the observed data \cite{rabiner1989tutorial}. The concept of hidden Markov models has been successfully employed in a variety of areas such as speech \cite{kayte2015hidden,muhammad2018speech,mustafa2019comparative}, gesture \cite{presti2014gesture,sagayam2018abc,sagayam2019probabilistic}, and text recognition \cite{du2016deep,wang2018comprehensive,shen2020novel}, bioinformatics \cite{marco2017multi,manogaran2018machine,benelli2021charting,emdadi2021auto,porter2021profile,tang2021quantifying}, signal processing \cite{zhou2016detection,yuwono2016automatic,cai2018energy}%, climatology, seismology
, and finance \cite{cao2014adaptive,nguyen2018hidden,zhang2019high,zheng2021regime}.

Among the recent financial applications, hidden Markov models have been used to understand the behavior of the decentralized, loosely regulated \cite{la2021doge},
and highly speculative \cite{malladi2021time} market of cryptocurrencies. From these studies, Giudici and Abu Hashish (2020) \cite{giudici2020hidden} investigated how the prices of Bitcoin \cite{nakamoto2008bitcoin} switch between “bear,” “stable,” and “bull” regimes. Koki et al. (2022) \cite{koki2022exploring} modeled the return series of the three highest capitalization cryptocurrencies, Bitcoin, Ether\cite{buterin2013ethereum} and Ripple \cite{schwartz2014ripple}. Similarly to the results of Giudici and Abu Hashish (2020) \cite{giudici2020hidden}, they found that the hidden Markov framework distinguished "bear", "stable" and "bull" regimes for the Bitcoin series, while, in the case of the Ether and Ripple, %series
it separated periods with different profit and risk magnitudes. Finally, Kim et al. (2021) \cite{kim2021dynamics} used the hidden Markov model to examine how cryptocurrency markets behave and react to social sentiment under various regimes.

Similar to the behavior of the crypto market, increasing attention has been paid to detecting and understanding its anomalies. Some of these works were focused on the anomalies of the underlying blockchain transaction network \cite{monamo2016unsupervised,dixon2019blockchain,akcora2020bitcoinheist,li2020dissecting,hassan2021anomaly,ofori2021topological}. Others employed social media data \cite{phillips2017predicting,victor2019cryptocurrency,nizzoli2020charting,mirtaheri2021identifying,nghiem2021detecting} or directly investigated the price % and volume
anomalies of various cryptocurrencies \cite{kamps2018moon}. A common cause of these anomalies is the so-called pump-and-dump price manipulation, in which scammers lure traders to buy a cryptocurrency at an artificially inflated price (pump), then quickly sell their previous holdings to profit (dump) \cite{nghiem2021detecting}. According to Li et al. (2021)\cite{li2021cryptocurrency}, pump-and-dump activities lead to short-term bubbles featuring dramatic increases in prices, volume, and volatility. Prices peak within minutes, followed by quick reversals. In contrast with Li et al. (2021)\cite{li2021cryptocurrency}, Hamrick et al. (2021)\cite{hamrick2021analyzing} investigated long-term (so-called target-based) pump-and-dump schemes, in which pump signals do not reach a buy target until days or even weeks.

In this paper, (1) we present a new method that is able to find linear laws governing the time evolution of Markov chains, then (2) we apply this method for anomaly detection in Bitcoin prices. To this end, we first derive the linear laws of Markov chains by using the time embedding of their (categorical) autocorrelation function. At this point, we rely on Jakovác's (2020, 2021)\cite{jakovac2020,jakovac2021time} ideas on linear laws as well as the time-delay embedding method proposed by Takens (1981)\cite{takens1981dynamical} and used in various fields, such as dimensional causality analysis\cite{benko2018exact,benkHo2019inferring,benkHo2019causal,zlatniczki2021relaxation,benkHo2022manifold} and anomaly detection\cite{benkHo2022model}. As a next step, we generate a binary series from the first difference of Bitcoin exchange rate (against the United States Dollar). Finally, we identify the minimum number of parameters describing the linear laws of this series through stepped time windows. Based on our results, linear laws typically became more complex (containing an additional third parameter that indicates hidden Markov property) in two periods: before the crash of cryptocurrency markets inducted by the COVID-19 pandemic (12 March 2020), and before the record-breaking surge in the price of Bitcoin (Q4 2020 -- Q1 2021). In addition, the locally high values of this third parameter are often related to short-term price peaks, which suggests price manipulation.

The paper is organized as follows. The linear laws of Markov chains are formulated in the Methods section. This is followed by the Results and Discussions section, which first demonstrates how linear laws are able to indicate the hidden Markov property of a series. Then, it presents and discusses the results of anomaly detection in Bitcoin prices. The paper ends with the Conclusions and Future Work section, which provides the conclusions and suggests future research directions.
%Finally, Conclusions and Future Work section provides the conclusions and suggests future research directions.

\section*{Methods}
\label{sec:methods}

\subsection*{Finding linear laws in Markov chains}

Markov chains are random processes, where the next element depends only on the present one, and not on the previous ones. Let us denote by $P((n,x)|C)$ the conditional probability to find in the $n$-th step $x\in V$, where $V$ is a vector space, and $C$ is some condition. In a causal process, the condition can be all the history, but in Markov processes it simplifies:
\begin{equation}
    P((n+1,x) | (n, y),\dots, (0, y_0)) = P((n+1,x) | (n,y)) \equiv T^{(n)}_{xy}.
\end{equation}
This $T^{(n)}_{xy}$ can be treated as a matrix if the range in $V$ is discrete. In the following we assume that there is no $n$ dependence (time translation symmetry), and we denote $T^{(n)}_{xy}\equiv T_{xy}$. The $T$ matrix is used to call the transfer matrix. $T_{xy}$ means the probability of the transfer $y\to x$. 

Since the elements of $T$ are probabilities, $0\le T_{xy}\le1$ must be true. Moreover, from the state $y$ we arrive to some state with probability one, thus we find:
\begin{equation}
    \sum_x T_{xy} = 1,
\end{equation}
the sum of each column is one. This means that the vector $v = (1,1,\dots,1)$ is a left eigenvector with eigenvalue one, $vT=v$. Because the spectrum is the same for left and right eigenvectors, there is a $\lambda=1$ eigenvalue in the spectrum.

If we know the probability distribution after the $n$-th step to be $P((n,y)) = \bar P^{(n)}_y$, then the probability distribution after the $(n+1)$-th step is:
\begin{equation}
    \bar P^{(n+1)}_z =P((n+1,z)) = \sum_{y\in V} P((n+1,z)|(n,y)) P((n,y)) = (T\bar P^{(n)})_z.
\end{equation}
This follows, after generalization:
\begin{equation}
    \bar P^{(m)} = T^{m-n} \bar P^{(n)}.
\end{equation}
Equilibrium distribution, denoted simply by $\bar P$ is invariant under the time translation, i.e.:
\begin{equation}
    \bar P = T \bar P,
\end{equation}
it is the right eigenvector of the transfer matrix with $\lambda=1$ eigenvalue.

With the help of the transfer matrix we can calculate the expected value of any function as (assume $n<n_1<\dots < n_k$):
\begin{eqnarray}
    && \left\langle f(x_n, x_{n_1}, x_{n_2}\dots x_{n_k})\right\rangle =\sum_{y_0,\dots, y_k\in V} f(y_0, \dots ,y_k) \times \nonumber\\
    && \times P((n,y_0))P((n_1,y_1)|(n,y_0))\dots P((n_k,y_k)|((n_{k-1},y_{k-1}))=\nonumber\\
    && = \sum_{y_0,\dots, y_k\in V} f(y_0, \dots ,y_k) (T^{n_k-n_{k-1}})_{y_ky_{k-1}}\dots (T^{n_1-n})_{y_1y_0}  \bar P^{(n)}_{y_0}.
\end{eqnarray}
We remark that this formula in the continuous case is called path integral. 

In particular, we may consider the expected value of a function of two variables:
\begin{equation}
    C^{(n)}_k[f] = \left\langle f(x_n, x_{n+k})\right\rangle = \sum_{y,z\in V} f(y ,z) \bar P^{(n)}_y (T^k)_{zy} = \mathrm{Tr}\; F^{(n)} T^k,
\label{eq:autocovariance}
\end{equation}
where $\mathrm{Tr}$ means trace and the $F$ matrix reads:
\begin{equation}
    F^{(n)}_{yz} = f(y,z) \bar P^{(n)}_y.
\end{equation}
As we see, for the two-variable case, the $n$ and $k$ dependence factorizes. We will consider long-time behavior, where $\bar P^{(n)}=\bar P$ is the equilibrium distribution. The result is independent of $n$, so we may omit this variable.
%The result is the $n$ independent, and we may omit this variable.

After we diagonalized the transfer matrix:
\begin{equation}
    T = U \Lambda U^{-1},\qquad \Lambda_{yy'} = \lambda_y\delta_{yy'}.
\end{equation}
We find for $C_k[f]$:
\begin{equation}
    C_k[f] = \mathrm{Tr} F U \Lambda^k U^{-1} = \sum_{y\in V} \lambda_y^k (U^{-1}FU)_{yy} = \sum_{y\in V} c_y \lambda_y^k,
\end{equation}
where:
\begin{equation}
    c_y=(U^{-1}FU)_{yy}.
\end{equation}
This means that $C_k[f]$ is in general a sum of geometric series, and the number of terms is equal to the number of possible states.

For this special class of series we may look for a linear law %\cite{jakovac2021time,jakovac201013482}
in the form:
\begin{equation}
    \sum_a w_a C_{k+a}[f] = 0,\quad\forall k.
\end{equation}
This is possible if:
\begin{equation}
    0 = \sum_a w_a \sum_{y\in V} c_y \lambda_y^{k+a} = \lambda^k\sum_{y\in V} c_y (\sum_a w_a \lambda_y^a). 
\end{equation}
Therefore we need:
\begin{equation}
    \sum_a w_a \lambda_y^a = 0,
\label{eq:linlaw}
\end{equation}
i.e., we seek those polynomials, whose roots are the eigenvalues of the transfer matrix. But we know such a polynomial, the equation which determines the eigenvalues, i.e., the characteristic equation for the transfer matrix:
%i.e. we seek those polynomials, whose roots are the eigenvalues of the transfer matrix. \textbf{The coefficients of these polynomials form the components of the linear law vector. Such polynomials determine the equation which....} But we know such a polynomial, the equation which determines the eigenvalues, i.e. the characteristic equation for the transfer matrix:
\begin{equation}
    \det(T-\lambda ) = 0.
\label{eq:characteristic_equation}
\end{equation}

This means that the expected value of any function of two variables, $C_k[f]$, satisfies a linear law for Markov processes. The coefficients of the linear law (defined by Eq. (\ref{eq:linlaw})) are equal to the coefficients of the characteristic polynomial of the transfer matrix, while its roots are the eigenvalues of the transfer matrix. This means that by inspecting the embedding of $C_k[f]$ function we can also determine the eigenvalues of the transfer matrix.

We emphasize that the above conclusion is true for arbitrary function of two variables. In particular we may use $f(x,y) = f(x_1,y_1)$, where $x_1$ is the first component of the $x\in V$ vector. Put another way, if we observe only a subprocess, we can determine the eigenvalues of the transfer matrix of the complete process.

\subsection*{Case of binary Markov chains}

The simplest case is the binary Markov chain, where $V=\{0,1\}$. The transfer matrix is a $2\times2$ matrix.
Since the sum of all columns is one, thus it contains two parameters:
\begin{equation}
    \label{eq:binaryt}
    T = \left(
    \begin{array}{cc}
        1-p & q \\
        p & 1-q \\
    \end{array}
    \right).
\end{equation}
Here $p$ is the probability of $0\to1$, $q$ is the probability of $1\to0$. The equilibrium distribution:
\begin{equation}
    \bar P = \frac1{p+q}\left(
    \begin{array}{c}
        q \\
        p \\ 
    \end{array}
    \right).    
\end{equation}
The characteristic polynomial of $T$ is:
\begin{eqnarray}
    && 0 = \left|
    \begin{array}{cc}
        1-p-\lambda & q \\
        p & 1-q-\lambda \\
    \end{array}
    \right| = (1-p-\lambda)(1-q-\lambda) - pq = \nonumber\\
    && = \lambda^2 - (2-p-q)\lambda + 1-p-q = (\lambda-1)(\lambda+p+q-1).
\end{eqnarray}
The two eigenvalues are $\lambda =1$ and $\lambda = 1-p-q$. Both are real, but not necessarily positive. The corresponding right eigenvectors are:
\begin{equation}
    U =  \left(\begin{array}{rr}
         r & -1\\
         1 & 1\\ 
    \end{array}\right),\qquad r =\frac qp.
\end{equation}
The left eigenvectors are:
\begin{equation}
    U^{-1} = \frac1{1+r}\left(\begin{array}{rr}
         1 & 1\\
         -1 & r\\ 
    \end{array}\right).
\end{equation}
We find:
\begin{equation}
    U^{-1}TU = \Lambda = \left(\begin{array}{cc}
         1 & 0\\
         0 & 1-p-q\\ 
    \end{array}\right),\qquad U\Lambda U^{-1} = T.
\end{equation}
As an example we may consider the $f(x,y) = \delta_{xy}$ function, i.e., we count the cases where $x=y$. This is the most appropriate choice for categorical variables. We find:
\begin{equation}
    F_{xy} = \delta_{xy} \bar P_y,\qquad c_y = \sum_{y'\in V} U^{-1}_{yy'} \bar P_{y'} U_{y'y} =\frac1{(1+r)^2} \left(\begin{array}{c}
         1+r^2  \\
         2r
    \end{array}\right),
\end{equation}
therefore:
\begin{equation}
    C_k[f] = \frac{1 + r^2 + 2r(1-p-q)^k}{(1+r)^2}.
\end{equation}
The characteristic polynomial of the transfer matrix is:
\begin{equation}
    \det(T-\lambda) = (1-\lambda)(1-p-q-\lambda) = = \lambda^2 - (2-p-q)\lambda + 1-p-q.
\end{equation}
Thus, as it is easy to check also directly, our $C_k[f]$ expression satisfies the linear law:
\begin{equation}
    C_{k+2} - (2-p-q) C_{k+1} + (1-p-q)C_k = 0.
\end{equation}

In a real system we observe $C_k$, and produce the matrix:
\begin{equation}
    F_{ka} = C_{k+a}.
\label{eq:feqc}
\end{equation}
We seek $w_a$ that satisfies:
\begin{equation}
    0 = \sum_a w_a C_{k+a} = (F\cdot w)_k.
\end{equation}
Multiplying this with $F^{T}$ we get:
\begin{equation}
    0 = \sum_{a,k} F_{ck} F_{ka} w_a  = \sum_a \left( F^{T} F \right)_{ca} w_{a}.
\label{eq:ftf}
\end{equation}
This means exactly that $F^T F$, which is a $3\times3$ matrix, has a zero eigenvalue. Then $w$ is the corresponding eigenvector.

We remark that the 2-state Markov chain can be written in a Langevin equation form as:
\begin{equation}
    x_{n+1} = \mathop{\mbox{int}}(1+p-\xi + (1-p-q)x_n),
\end{equation}
where $\xi \in [0,1]$ with uniform distribution, and "int" means integer part. As it is easy to check, the above formula yields $x_{n+1}\in\{0,1\}$ if it was true for $x_n$, and the transfer matrix is exactly the same as shown in the Eq. \eqref{eq:binaryt}.

\section*{Results and Discussion}

%\subsection*{Complexity of Linear Laws}

\subsection*{Detection of Hidden Markov Property}
%In this subsection, we show how to determine the necessary and sufficient number of parameters to identify the linear laws of simulated Markov series in practice. First, 
%In this subsection, we show how the proposed method helps to determine the necessary and sufficient number of parameters (as well as the linear laws itself) for simulated Markov series.
%In this subsection, we show how the proposed method helps to identify the complexity of a Markov series. 

\begin{figure}[ht]
     \centering
     \begin{subfigure}[ht]{0.3\textwidth}
        \centering
         \caption{For $x$ series}
         \includegraphics[width=\textwidth]{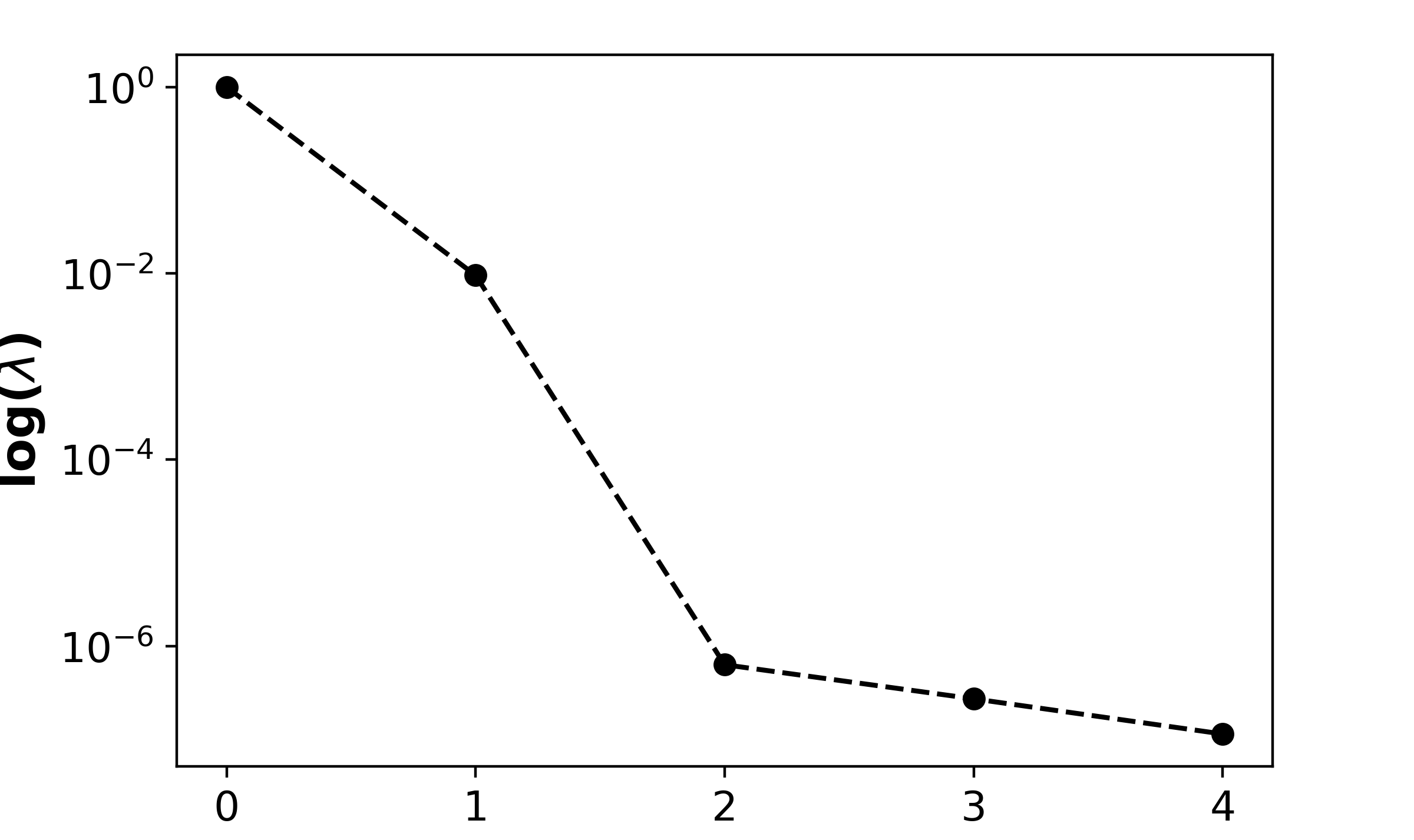}
     \end{subfigure}
     \begin{subfigure}[ht]{0.3\textwidth}
         \centering
         \caption{For $y$ series}
         \includegraphics[width=\textwidth]{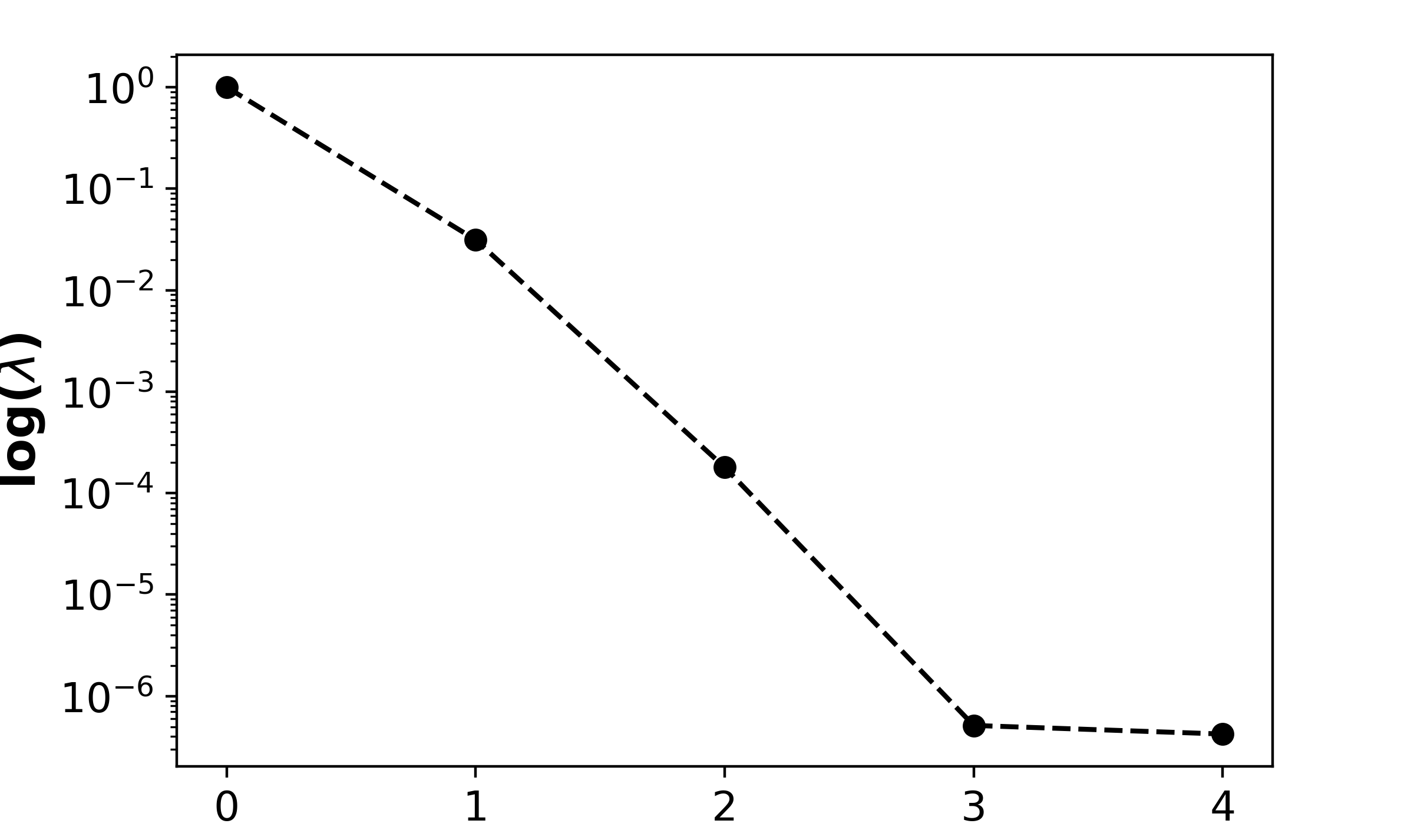}
     \end{subfigure}
     \begin{subfigure}[ht]{0.3\textwidth}
         \centering
         \caption{For $z$ series}
         \includegraphics[width=\textwidth]{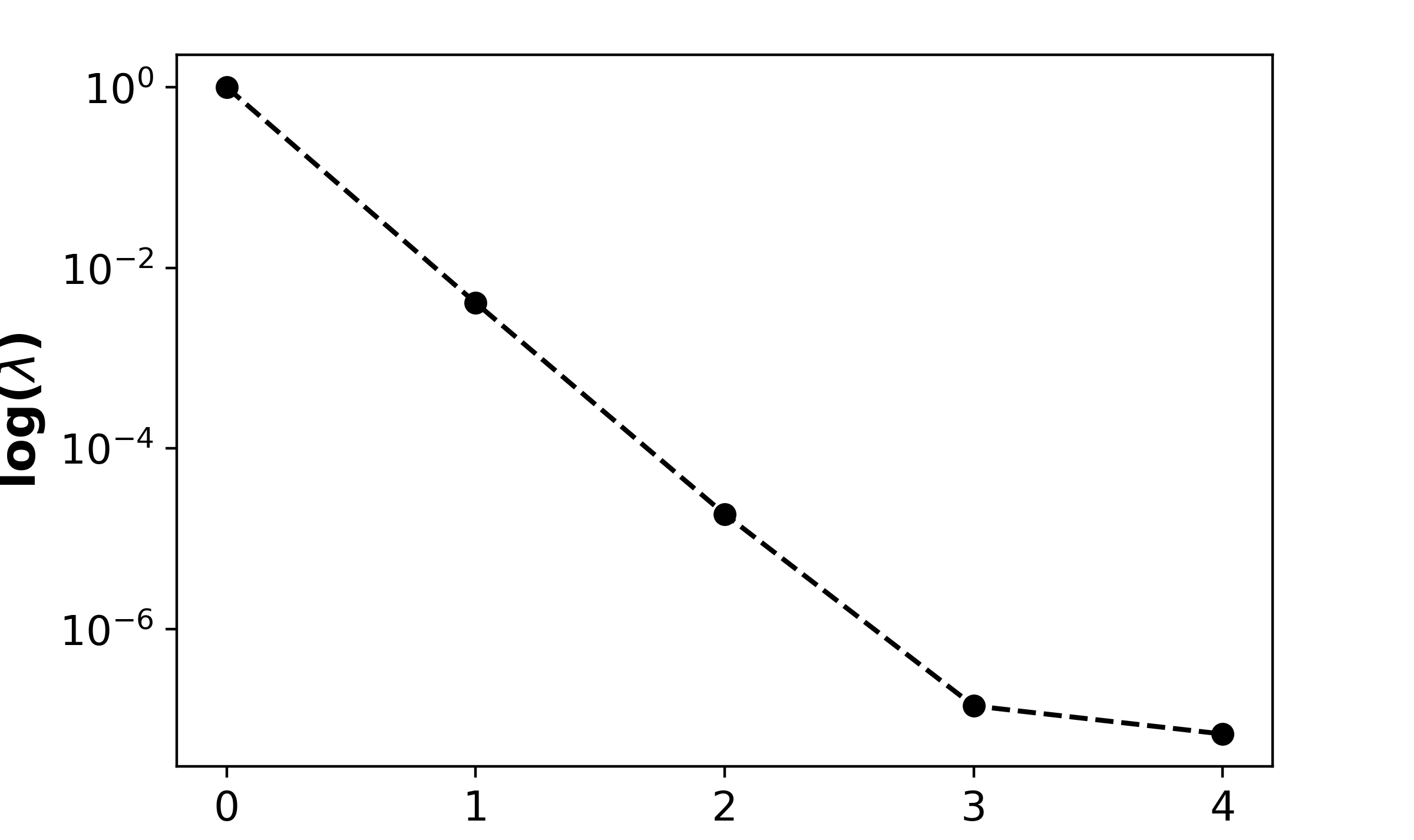}
     \end{subfigure}
    \caption{Eigenvalues ($\lambda$) of the $F^{T}F$ matrix in descending order %for series (a) $y_n$ and (b) $x_n$
($k = 20$ and $n = 5$ were applied and the eigenvalues were divided by the highest eigenvalue).}
    \label{fig:eigenvalues}
\end{figure}

In this subsection, we demonstrate how the proposed method reveals the hidden Markov property of a simulated process, through the example of three simulated series. First, %let us consider
a $1\,000\,000$-length binary Markov chain ($x$) is simulated by using the following (randomly generated) transfer matrix ($T_{x}$):

\begin{equation}
    \label{eq:BinaryTmatrix}
    T_{x} = \left(
    \begin{array}{cc}
    0.3768 & 0.6335 \\
    0.6232 & 0.3665
    \end{array}
    \right),
\end{equation}

\noindent
and $x_0 = 0 \in V_x$ as initial state. Then, we generate an other $1\,000\,000$-length Markov process ($y$) with $V_y = \{00,10,01,11\}$ states based on the following (random) transfer matrix ($T_{y}$):
%Then, an other $1\,000\,000$-length Markov process ($y$) with $V_y=\{00,10,01,11\}$ states is simulated by using the following (randomly generated) transfer matrix ($T_{y}$):

\begin{equation}
    \label{eq:BinaryTmatrix1}
    T_{y} = \left(
    \begin{array}{cccc}
        0.1323 & 0.3055 & 0.2635 & 0.1005 \\
        0.4632 & 0.1256 & 0.0126 & 0.3680 \\
        0.3622 & 0.3303 & 0.6189 & 0.1519 \\
        0.0423 & 0.2386 & 0.1049 & 0.3796
    \end{array}
    \right),
\end{equation}

\noindent
and $y_0 = 01 \in V_y$ initial state. %Similarly to Eq. (\ref{eq:twobits}),
The left bit of this process forms a separate binary series ($z$) with $z_0 = 0 \in V_z$ initial state. Applying Eqs. (\ref{eq:binaryt}) -- (\ref{eq:ftf}) to $x$, $y$, and $z$, the eigenvalues of the given $F^{T}F$ matrices are determined. These eigenvalues (contained by $\lambda_x$, $\lambda_y$, and $\lambda_z$) are illustrated in Fig. \ref{fig:eigenvalues}.

The difference between $\lambda_x$ and $\lambda_z$ suggests that although both $x$ and $z$ are binary processes ($V_x = V_z = \{0,1\}$), unlike $x$, $z$ cannot be characterized by two parameters (see $p$ and $q$ in Eq. (\ref{eq:binaryt})). In other words, it is only the part of the $y$ Markov process of similar complexity (cf. $\lambda_y$ and $\lambda_z$ in Fig. \ref{fig:eigenvalues}). This phenomenon refers to the hidden Markov property of the $z$ series.

\subsection*{Anomaly detection in Bitcoin prices}

We first generate a binary series ($\rho_t$) from the first difference of Bitcoin exchange rate (against the United States Dollar) as follows:

 \begin{equation}
    \rho_t =
    \begin{cases}
      1, & \text{if}\ \quad \pi_{t} - \pi_{t-1} \geq 0 \\
      0, & \text{otherwise},
    \end{cases}
  \end{equation}

\noindent
where $\pi_{t}$ is the closing price of Bitcoin in time $t$. Similar to Phaladisailoed and Numnonda (2018)\cite{phaladisailoed2018machine}, Kavitha et al. (2020)\cite{kavitha2020performance}, and Passalis et al. (2021)\cite{passalis2021forecasting}, we use the 1-minute interval price data of Bitstamp (collected from: \url{https://www.cryptodatadownload.com/}, retrieved: 9 January 2022) %(\url{https://www.bitstamp.net/}, retrieved: 10 December 2021)
during the calculations. The employed dataset %dataset we employed 
contains $2\,641\,395$ 
%$4\,857\,377$
price data (at a missing value rate of $1.49\%$)
%at a missing value rate of $25.7\%$) 
%$25.6\%$
between 1 January 2017 00:01:00 GMT and 9 January 2022 07:16:00 GMT ($t\in\small\{1,2,...,2\,627\,500\small\}$).
%3\,613\,769\small\}$).
%31 March 2021 00:00:00 GMT 
%($t\in\small\{1,2,...,3\,613\,769\small\}$).

\begin{figure}[ht]
    \centering
    \includegraphics[width=0.9\textwidth]{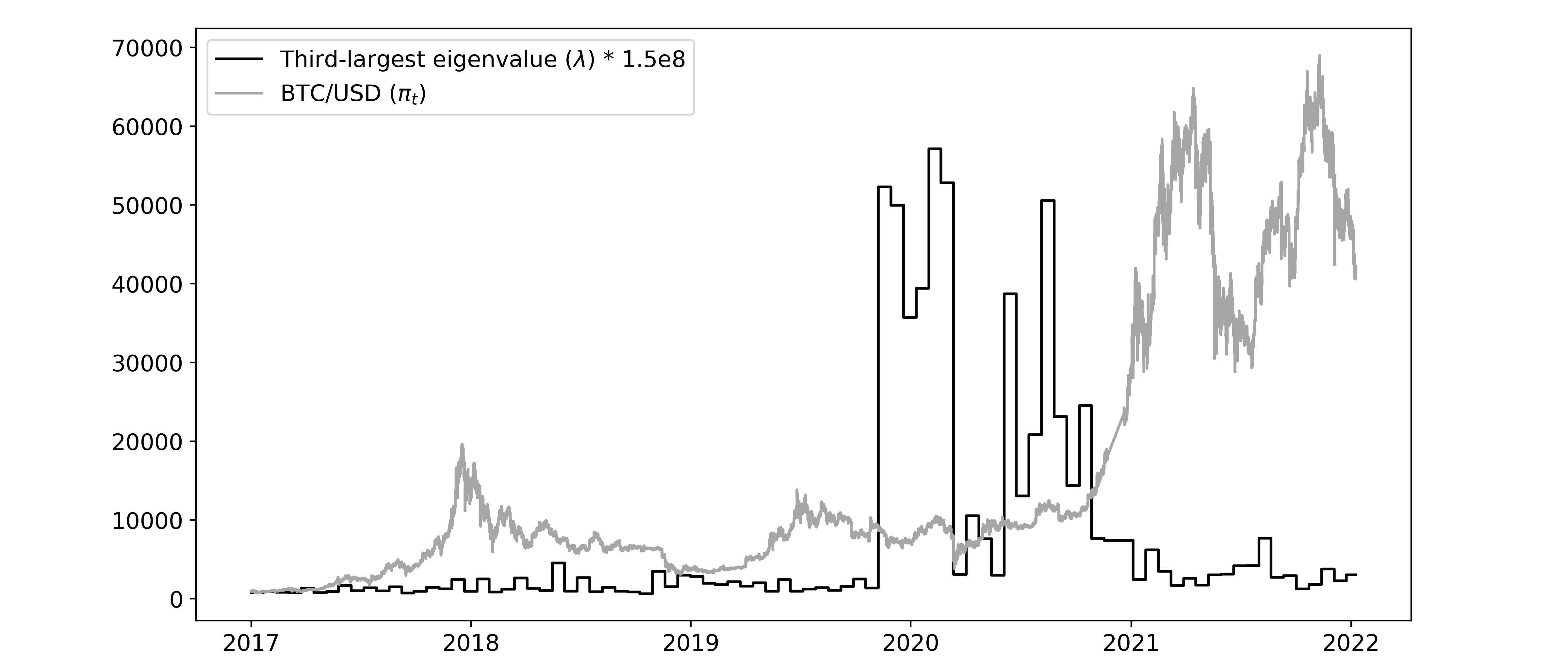}
    \caption{Bitcoin exchange rate in United States Dollar (BTC/USD) and third-largest eigenvalues identified through 30\,000-width stepped time windows ($k = 20$ and $n = 5$ were applied and and the eigenvalues were divided by the highest eigenvalue of the given time window).}
    \label{fig:btcusd}
\end{figure}

As a next step, the studied time horizon is divided into 30\,000-width time windows (i.e., the first time window contains price data from $t=1$ to $t=30\,000$). Then, by applying Eqs. (\ref{eq:binaryt}) -- (\ref{eq:ftf}) to $\rho_t$, the eigenvalues of $F^{T}F$ matrices are determined for each time window. As discussed in the previous section, the magnitude of the third-largest eigenvalue can indicate that two parameters (see $p$ and $q$ in Eq. (\ref{eq:binaryt})) are not sufficient to describe the observed binary series. Following this logic, Fig. \ref{fig:btcusd} illustrates the third-largest eigenvalues determined in each time window.

Based on the results, the examined eigenvalues are remarkably high in two periods: before the crash of cryptocurrency markets inducted by the COVID-19 pandemic (12 March 2020), and before the record-breaking surge in the price of Bitcoin (Q4 2020 -- Q1 2021). Although it is difficult to determine the reasons for this phenomenon, it is striking that most of the locally high eigenvalues are related to a short-term price peak. Thus, a possible explanation is that intensive pump-and-dump activities artificially manipulate Bitcoin prices and permanently change its evolution over time. %manipulate Bitcoin prices, permanently changing its evolution over time.
This is also supported by the fact that the anomaly disappeared immediately after the collapse of Bitcoin prices (12 March 2020), and reappeared only after the uncertainty caused by COVID-19 decreased (June 2020).

%reappeared only after reducing the uncertainty following the announcement of COVID-19 pandemic (June 2020).

%reappeared only after the uncertainty caused by COVID-19 decreased (June 2020).

\section*{Conclusions and Future Work}

This paper presented a new method that is able to find linear laws governing the time evolution of Markov chains by using the time embedding of their (categorical) autocorrelation function. As an application, we investigated the linear laws of the 1-minute interval price data of Bitcoin between 1 January 2017 00:01:00 GMT and 9 January 2022 07:16:00 GMT. To this, first, a binary Markov series was generated from the first difference of the exchange rate of Bitcoin (against the United States Dollar). Then, the minimum number of parameters describing the linear laws of this series was identified through stepped time windows.

Our investigation demonstrated promising results on the detection of Bitcoin anomalies. Based on our findings, linear laws typically became more complex (containing an additional third parameter that indicates hidden Markov property) in two periods: before the crash of cryptocurrency markets inducted by the COVID-19 pandemic (12 March 2020), and before the record-breaking surge in the price of Bitcoin (Q4 2020 -- Q1 2021). In addition, the locally high values of this third parameter are often related to short-term price peaks, which suggests price manipulation.

In our future work, we will examine the linear laws in exchange rates of other cryptocurrencies, as well as stock prices and the prices of oil and other commodities. Although due to the theory of efficient market hypothesis\cite{fama1970}, financial data typically fit well with the concept of Markov chains, we also plan to investigate electroencephalograms (EEGs) record neural activity to detect epilepsy.

%\section*{Data Availability}

%The main data supporting the findings are provided with this paper. All the raw data generated in this study are available from the corresponding author upon reasonable request.

\bibliography{sample}

%For data citations of datasets uploaded to e.g. \emph{figshare}, please use the \verb|howpublished| option in the bib entry to specify the platform and the link, as in the \verb|Hao:gidmaps:2014| example in the sample bibliography file.

\section*{Acknowledgements}

The authors would like to thank Andr\'as Telcs (Wigner Research Centre for Physics, Budapest) and Zolt\'an Somogyv\'ari (Wigner Research Centre for Physics, Budapest) for their valuable comments and suggestions. The authors thank the support of E\"otv\"os Lor\'and Research Network. The research was supported by the Ministry of Innovation and Technology NRDI Office within the framework of the MILAB Artificial Intelligence National Laboratory Program. A.J. had a support from the Hungarian Research Fund NKFIH (OTKA) under contract No. K123815.

\section*{Author contributions statement}

A.J., P.P., and M.T.K. conceptualized the work and contributed to the writing and editing of the manuscript. M.T.K. acquired the data and conducted the analysis. A.J. supervised the research.
%A.J., P.P., and M.T.K. conceptualized the work and contributed to the writing and editing of the manuscript. M.T.K. acquired the data and conducted the analysis and A.J. supervised the research.
%A.J., P.P., and M.T.K. contributed to the writing and editing of the manuscript.
%A.J. supervised the research.

\section*{Competing interests}

The authors declare no competing interests.

%\section*{Additional information}
%To include, in this order: \textbf{Accession codes} (where applicable); \textbf{Competing interests} (mandatory statement). 
%The corresponding author is responsible for submitting a \href{http://www.nature.com/srep/policies/index.html#competing}{competing interests statement} \hl{on behalf of all authors of the paper. This statement must be included in the submitted article file.}

\end{document}